\documentstyle[12pt,epsfig]{article}
\begin{document}
\thispagestyle{empty}
\setcounter{page}0

\vspace*{1.5 cm}
\begin{center}
{\LARGE{\bf More General BBN Constraints
\ \\
on Neutrino Oscillations Parameters}
\ \\
\Large{\bf Relaxed or Strengthened}}

\vspace*{0.4cm}
Daniela Kirilova$^{\dagger}$\\[0.3cm]
{\it $^\dagger$Institute of Astronomy, BAS, Sofia~\footnote{Regular
Associate of Abdus Salam ICTP}}
\end{center}
\vspace*{0.2cm}
\begin{abstract}

I discuss BBN with  nonequilibrium $\nu_e\leftrightarrow \nu_s$  oscillations
in the more general case
of non-zero population of $\nu_s$ before oscillations
 $\delta N_s\ne0$.
I calculate $^4$He primordial
production  $Y_p(\delta N_s$) in the
presence of  $\nu_e\leftrightarrow \nu_s$  oscillations for
different initial
populations of the sterile neutrino state $0\le\delta N_s\le1$ and the full range
of oscillation parameters.

{\it Non-zero} $\delta N_s$ {\it has two-fold effect} on  $^4$He: (i) it enhances the
energy density  and hence increases the
cosmic expansion rate, leading to   $Y_p$
{\it overproduction}  and
(ii) it suppresses the kinetic effects of oscillations on BBN, namely
 the effects on pre-BBN nucleon kinetics caused by the  $\nu_e$ energy
spectrum
distortion and the  $\nu_e-\bar{\nu_e}$ asymmetry generation by
oscillations, leading to {\it decreased} $Y_p$ {\it overproduction}.
Depending on oscillation parameters one or the other effect may dominate,
causing  correspondingly either   a relaxation of the cosmological constraints
 or  their strengthening
with the increase of $\delta N_s$ .

I calculate  more general BBN constraints on
 $\nu_e\leftrightarrow \nu_s$  oscillation  parameters, corresponding to $3\%$ $Y_p$
overproduction,   for
different  initial populations of the sterile state.
Previous BBN constraints were derived assuming
empty  sterile state before oscillations.
The cosmological constraints for that case
strengthen with the increase of $\delta N_s$ value,
the change being more considerable for  nonresonant oscillations.

\end{abstract}

\vspace*{0.5cm}
\section{Introduction}

Big Bang Nucleosynthesis (BBN)  provides one of the most sensitive probes
of the physical conditions
of the early Universe and is often used to constrain neutrino
oscillations.

Active-sterile neutrino oscillations may considerably effect the early
Universe, and in particular BBN. They
are capable of  exciting additional light particles into
equilibrium, thus, affecting the expansion rate $H$, and  they may also
distort neutrino energy spectrum
 and generate neutrino-antineutrino
asymmetry,
thus, in case of
electron-sterile oscillations $\nu_e \leftrightarrow \nu_s$, influencing the weak interaction rates of $\nu_e$  and, hence, the
kinetics of nucleons at the pre-BBN epoch and the primordial nucleosynthesis.

Among the light elements produced primordially,  $^4$He is the most
abundantly produced, most precisely measured and
calculated~\cite{He,cuoco,cyburt,cyb,brux},
and usually it is the chosen element
for putting BBN constraints on oscillations.
Cosmological constraints on neutrino oscillations parameters, based on
BBN and $^4$He observational data,  have been
discussed in numerous
 publications, see for example refs.~\cite{d81,kir,bd,CC,nashi,NPS,DiBari,dol2003,Villante}, the
 review papers~\cite{dol,dubnastro,CESJ} and the references there in.

All available constraints were obtained assuming that the sterile state was
empty before oscillations  $\delta N_s=N_{\nu}-3=0$, i.e. that
it has been filled only due to
active-sterile mixing during oscillations. $N_{\nu}$ is the number of
neutrino species in equilibrium.

In general, however,  $\nu_s$ state may not be initially empty.
 There exist
numerous possibilities of producing $\nu_s$  before
 $\nu_e \leftrightarrow \nu_s$ oscillations: from
  GUT models,   large extra
dimensions models,  Manyfold Universe models,
 mirror matter models, etc.  They may be also produced in preceding
 $\nu_{\mu,\tau}\leftrightarrow \nu_s$ oscillations
 in  4-neutrino and 5-neutrino  mixing schemes.~\footnote{The
 analysis of experimental oscillations data  provides
  some constraints on the sterile neutrino impact in
the oscillations explaining atmospheric and solar neutrino anomalies~\cite{constraints}, and hence,
provides some indications about the possible values of $\delta N_s$, which could have been eventually  produced by
oscillations.} The degree of initial population of $\nu_s$,  $\delta N_s$,
 and its energy spectrum  depends on the
concrete model of $\nu_s$ production.

Non-zero initially $\delta N_s$ can influence the
dynamical and the kinetic effects of oscillations on BBN~\cite{MPD} and
change the cosmological limits on oscillations.
The aim of this work is to define how and to what extend  the
available cosmological constraints  on $\nu_e \leftrightarrow \nu_s$
oscillation parameters, obtained with the assumption $\delta N_s=0$,
are  changed  in the
more general case of initially non-zero sterile population
 $\delta N_s \ne 0$.

In the next section I discuss how  $\delta N_s
\ne 0$ influences oscillations effects on BBN and provide
qualitative considerations of the expected change of cosmological constraints.
In section 3 I present the results of the numerical analysis of
 $^4$He production in BBN with nonequilibrium $\nu_e\leftrightarrow \nu_s$ oscillations
and  initially  nonzero $\nu_s$ population $0< \delta N_s <1$ and  discuss  the generalized
 cosmological constraints on oscillation parameters. In the last section  the results and the assumptions of the model are resumed and  possibilities of relaxing the cosmological constraints are discussed.

\section{Generalizing BBN Constraints on Oscillations}

I discuss the generalization of BBN constraints on $\nu_e \leftrightarrow \nu_s$ neutrino oscillations
 parameters for the case of oscillations effective after electron neutrino decoupling.
 It is known that
 at lower mass differences, when
sterile neutrino production takes place after active neutrino
decoupling, due to $\nu_e\leftrightarrow \nu_s$
oscillations between initially empty $\nu_s$ and electron
neutrino,
 i.e. for $(\delta m^2/eV^2)\sin^42\theta<10^{-7}$,
the re-population of active neutrino becomes slow and
 kinetic
equilibrium may be strongly broken by $\nu_e \leftrightarrow \nu_s$ oscillations.
I will denote these oscillations with an initially  non-equilibrium sterile neutrino further on 'nonequilibrium oscillations' for short.
They may cause  considerable
deviations of the  $\nu_e$ energy spectrum from the
equilibrium Fermi-Dirac form~\cite{kir,nashi} (called
further 'spectrum distortion') and generate
neutrino-antineutrino asymmetry~\footnote{The  asymmetry effect at large mixings
and small mass differences,  discussed here, is subdominant~\cite{nashi} and leads to a slight
suppression of the spectrum distortion effect. This is in contrast to the case of
 generation of lepton asymmetry by oscillations at
relatively high $\delta m^2$  and small mixings~\cite{fv96}, where it may play the dominant role.},  thus
influencing the weak interaction
rates of  the processes governing pre-BBN nucleon kinetics
$\nu_e + p \leftrightarrow n + e^+$,
$\bar{\nu_e} + n \leftrightarrow  p + e$,  $n \leftrightarrow  p + e + \nu_e$ the nucleons freezing and correspondingly the  primordial elements production.~\footnote{The effect of distortions on the energy distributions of neutrinos caused by residual interactions of neutrinos after 2 MeV during electron-positron annihilations was not considered, because it leads to a negligibly small change in
$Y_p$.~\cite{serpico}}

For that oscillation case a numerical analysis of the evolution of the
neutrino and nucleons, using the kinetic
equations for the neutrino density matrix and neutrino number densities in momentum
space was provided to make a proper account for the neutrino spectrum distortion,
depletion and neutrino asymmetry growth at each momentum~\cite{nashi}.
The analysis of such oscillations,  accounting precisely
 for the kinetic effect,  allowed to put stringent
constraints on oscillations effective after active neutrino
decoupling~\cite{nashi, NPS, dubnastro}.

The analytical fits to the exact constraints, corresponding to $3\%$ $^4$He
overproduction,  are~\cite{nashi, NPS, dubnastro}:
\begin{eqnarray*}
\delta m^2 (\sin^22\vartheta)^4\le 1.5 \times10^{-9} {\rm eV}^2
   ~~~\delta m^2>0  \\
                 |\delta m^2| < 8.2\times 10^{-10} {\rm eV}^2
               ~~~~\delta m^2<0,~~ large~~ \vartheta,
\end{eqnarray*}

These constraints, as well as all existing in literature BBN constraints, concern the case of empty sterile neutrino state at the start of oscillations.

The presence of a  nonzero  $\delta N_s$  before oscillations exerts two types of effects on BBN~\cite{MPD}:

(i) It increases the energy  density by
$\delta \rho=7/8(T_{\nu}/T_{\gamma})^4 \delta N_s \rho_{\gamma}$, thus modifying the cosmic expansion rate
$H=\sqrt{8\pi\rho/3M_p^2}$, which  reflects into
higher
freezing temperature of the nucleons and overproduction of $^4$He~\cite{shvartsman}.
The dynamical effect of  initially present   $\delta N_s$ on primordial
$^4$He production  will be  denoted further on by  $\delta Y_d$.
Due to this effect  {\bf strengthening
of the cosmological bounds}  with respect to the ones calculated at
$\delta N_s=0$  should be expected.

In case of   $\nu_{\mu,\tau}\leftrightarrow \nu_s$ oscillations the dynamical effect  is the only
effect of  non-zero  $\delta N_s$ present before oscillations.
In the case of  $\nu_e\leftrightarrow \nu_s$ oscillations {\it with almost
equilibrium neutrino energy distribution}, i.e. oscillations taking place
before neutrino decoupling,  this
is the leading  effect as well.
This effect  can be accounted for simply  by adding the
initial  $\delta N_s$ value to the one produced in oscillations.
So, in both these cases the rescaling of
the existing constraints is rather straightforward.

(ii)
In the  nonequilibrium $\nu_e\leftrightarrow \nu_s$ oscillations
case, the presence of partially populated
$\nu_s$  suppresses
the oscillations effects on pre-BBN nucleons kinetics~\cite{MPD}.
Further on  this kinetic effect is denoted by $\delta Y_k^s$, while the kinetic effects of oscillations are denoted by $\delta Y_k$ (or in terms of the effective degrees of freedom  $\delta N_k$,  where  $\delta Y_k \sim 0.013\times \delta N_k$).

 Kinetic effects
are a result of the
generated energy spectrum distortion of $\nu_e$ in oscillations between active and sterile neutrino,
and to a smaller degree are due to the neutrino-antineutrino asymmetry,
generated by oscillations.~\footnote{Neutrino-antineutrino asymmetry
generated during the resonant transfer of neutrinos
  exerts back effect on  oscillating
neutrino.
Although its
value is too small to have a direct kinetic effect on the synthesis
of light elements, i.e.  $L<<0.01$,  it effects indirectly
BBN by suppressing oscillations at small mixing angles, leading to less
overproduction of He-4 compared to the case without the account of
asymmetry growth \cite{nashi}.}
Both are very sensitive to the degree of population of $\nu_s$ because the rate of oscillations is energy dependent. The spectrum distortion effect plays however the dominant role, so  it will be discussed in more detail further on, although we have accounted precisely for both.

The distortion of $\nu_e$ spectrum  leads both to a {\it depletion of
the active neutrino number densities} $N_{\nu}$:
$N_{\nu}\sim \int {\rm d}E E^2 n_{\nu}(E)$
and a decrease of the
$\Gamma_w$,  causing an earlier $n/p$-freezing and
an overproduction of $^4\!$He
yield. The spectrum distortion is the greatest in the
case the
sterile state is empty at the start of oscillations, $\delta
N_s=N_{\nu_e}/N_{\nu_s}=0$. It
decreases with the increase of the degree of population of the sterile
state at the onset of oscillations as illustrated
in
the following figure from ref.~\cite{MPD}.
\ \\

\hspace{2cm}\includegraphics[scale=0.4]{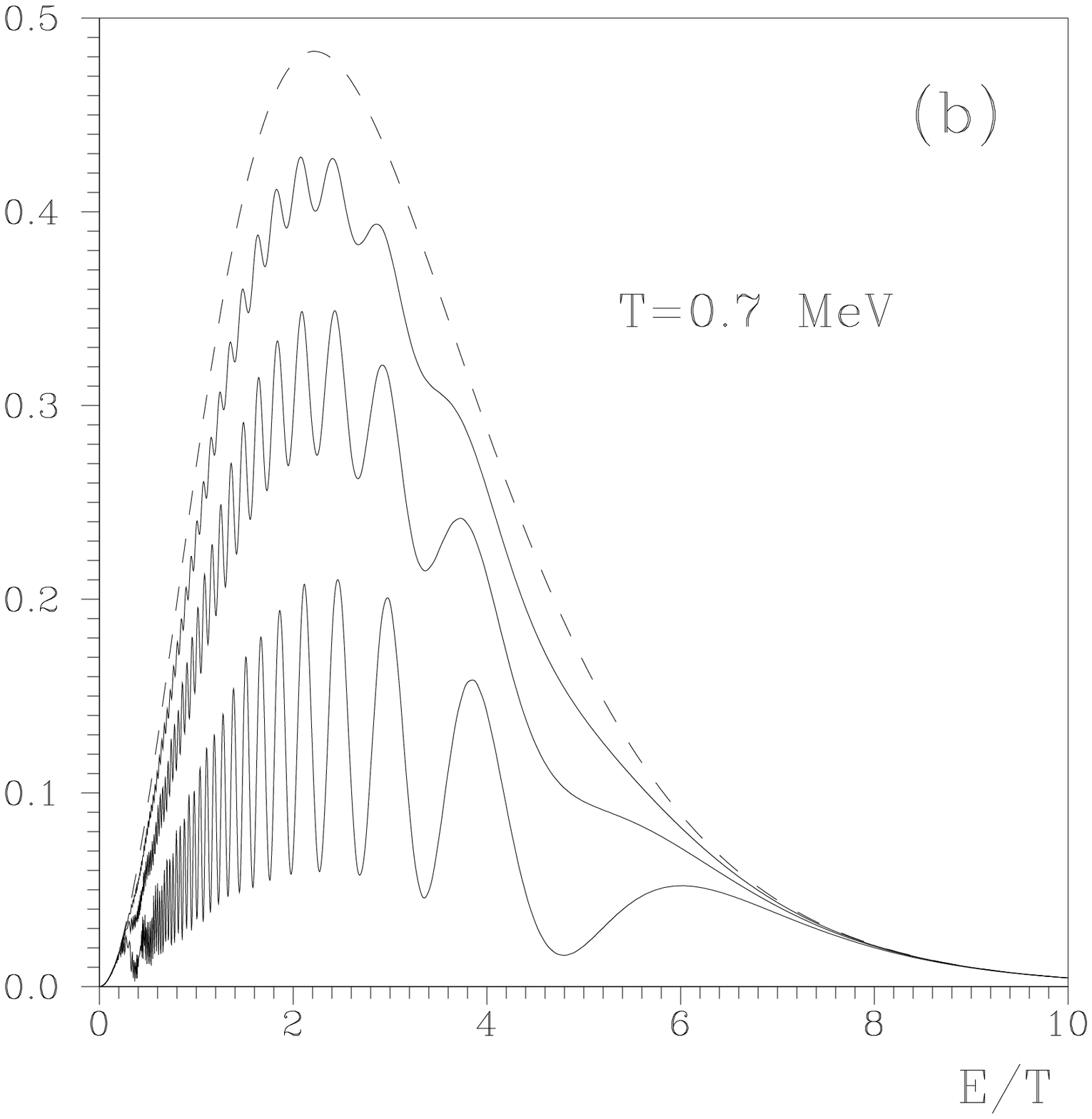}
\ \\
{\footnotesize Fig.1. The figure gives a snapshot of the spectrum distortion
of the electron neutrino energy spectrum $x^2\rho_{LL}(x)$, where
$x=E/T$  at a characteristic temperature  $0.7$ MeV,
caused by resonant oscillations with a mass difference
  $\delta m^2=10^{-7}$
eV$^2$ and mixing $\sin^22\vartheta=0.1$
 for
different degrees of initial sterile neutrino  population, namely $\delta N_s=0$
(lower curve), $\delta N_s=0.5$ and $\delta N_s=0.8$ (upper curve).
The
dashed curve gives the equilibrium spectrum for comparison.}

So, larger $\delta N_s$  leads to a decrease  of
the kinetic effects of oscillations and to a decrease of the
overproduction of $^4$He by oscillations with respect to the case of
initially zero  $\delta N_s$, i.e.  $\delta N_k<\delta N^0_{kin}$, where $\delta N^0_{kin}$ is the kinetic effect corresponding
to zero initial population of the sterile state, which presents in fact the  maximal kinetic effect
at a given set of oscillation parameters.~\footnote {For  $\nu_e\leftrightarrow \nu_s$  oscillations effective
after
$\nu_e$ decoupling, these kinetic effects $\delta N^0_k$   can be
considerable, as large as $\delta N^0_k=6$~\cite{ap}.}
 Due to  the decrease $\delta Y^s_k$ in the
overproduction of $Y_p$, caused by the  suppression of $\delta N_k$ by
the initially present  $\delta N_s \ne 0$,
 {\bf relaxation of the
cosmological bounds} with respect to the ones
corresponding to an initially empty  $\nu_s$ state should be expected.

Obviously, there is an  interplay between the two types of effects (i) and
(ii)
induced  by non-zero
$\delta N_s$. The total effect depends  on the concrete values of the
oscillation parameters and $\delta N_s$. {\it  The shift of the constraints may be
expected in either direction (relaxing
or constraining the existing constraints)}  and its analysis requires
numerical study, accounting precisely for $\nu_e$ energy spectrum distortion and the
suppression of the kinetic effects, which is essential.

Using the approximate empirical formula giving the dependence of $Y_p$
 production on the  two effects, obtained in ref.~\cite{MPD}:
$$
\delta Y_p = \delta Y_d + \delta Y^0_k + \delta Y^s_k \sim 0.013\times(\delta N_s +\delta N^0_{kin}-\delta N_s \times\delta N^0_{kin})
$$
it is possible to make some predictions  concerning the value and the sign of the $^4$He
overproduction and the direction of the shift of the BBN.
Here  $\delta Y_p=Y_p-Y_p^{stand}$,  $\delta Y_d=0.013 \times\delta N_s$,
$\delta Y^0_k=0.013\times\delta N^0_{kin}$, while
$\delta Y^s_k=-0.013 \times\delta N_s \times\delta N^0_{kin}$.

 {\bf First:  The total  effect of oscillations for initially   non-zero} $\delta N_s$ {\bf on $^4$He  production
is smaller than the   sum of
the energy density increase effect (i)} $\delta Y_d$ {\bf and the maximum kinetic effect
of oscillations (ii)}  $\delta Y^0_{k}$ corresponding to zero
 $\delta N_s$, due to the term  $\delta N_s \times\delta N^0_{kin}$ expressing
the decrease of  oscillations kinetic effect   with $\delta N_s$.

The results of the exact numerical study  of
the effects (i) and (ii) on $^4$He overproduction, illustrated in the following figures,
 confirm this
estimation. In  Fig.2a and Fig.2b  the contribution of the different effects on neutron-to-nucleon freezing ratio $X_n=n_n^f/n_{nuc}$  is presented. (The $Y_p$ production mainly depends on $X_n$,  $Y_p\sim X_n\times exp(-t/\tau_n)$, where $\tau_n$ is the neutron lifetime.)
 The dotted curve shows the kinetic effect dependence on
$\delta N_s$, the lower dashed curve gives the energy increase effect.
The total effect is presented by the solid curve, which although has different behavior in the two cases (decreasing or increasing), is situated
 considerably lower  than the uppermost long-dashed curve presenting  the sum of the
effects (i) and (ii).
\ \\

\hspace{2cm}\includegraphics[scale=0.4]{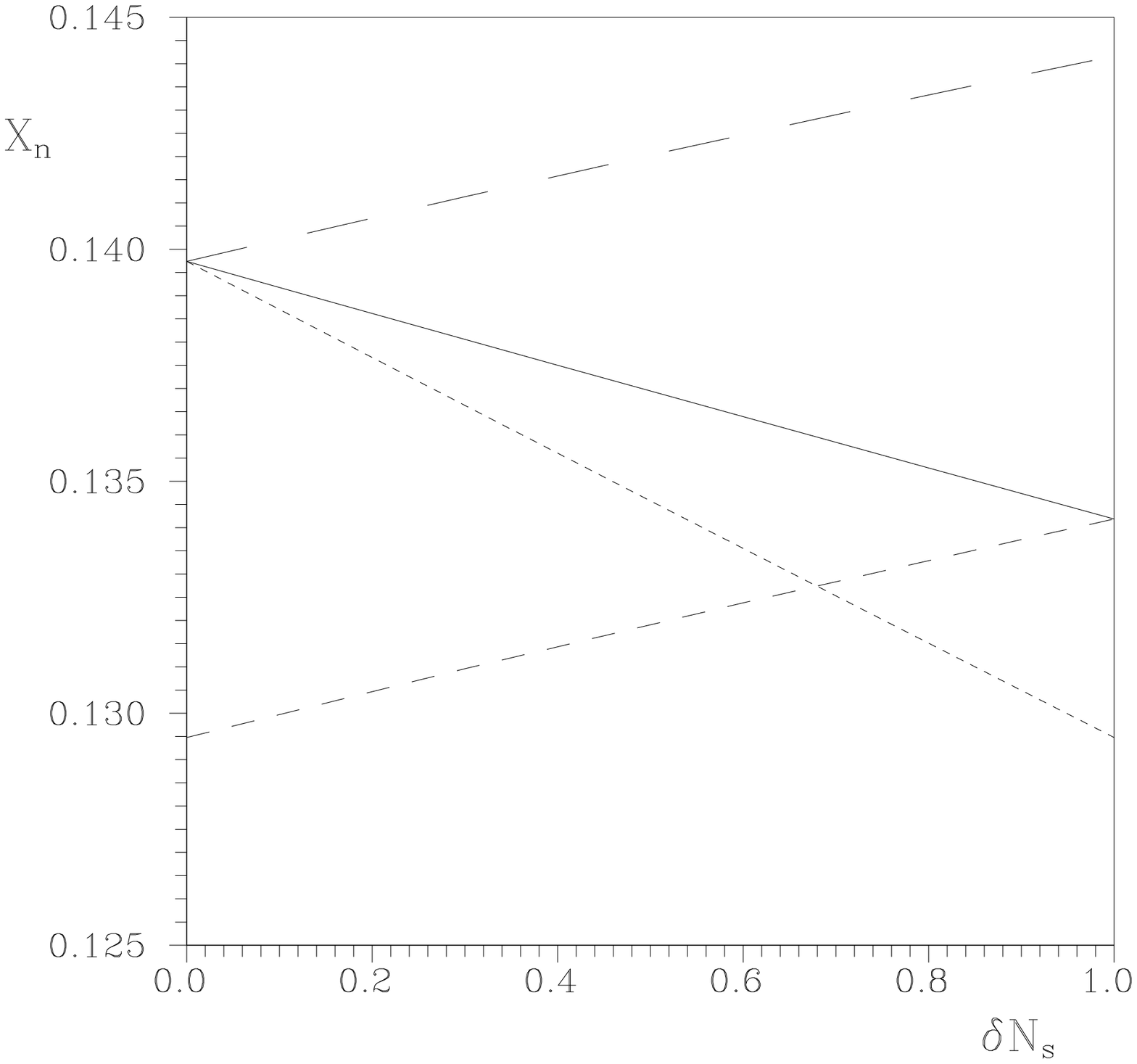}
\ \\
{\footnotesize Fig.2a. The  solid curve presents the frozen
neutron number density relative to
nucleons $X_n=n_n^f/n_{nuc}$ as a function  of
the sterile neutrino initial population, at $\delta m=\pm 10^{-8}$
eV$^2$, $\sin^2 2\theta=1$. The dotted curve presents
the
kinetic effect, while the lower dashed curve presents
energy
density increase effect. The uppermost long dashed curve corresponds to the
total effect when the decrease of the kinetic effect is not accounted
for,
as if   the initial
$\delta N_s$ were stored in a state not participating in oscillations with electron neutrino,
and the effects were simply additive.}

{\it Hence, the cosmological constraints
will be less stringent than in the case of simply  additive
 effects.}
 The latter case will have place
when   the enhancement of the energy density $\delta N_s$ is due to
other additional particles brought partially into equilibrium (like sterile neutrinos in
the muon or tau-sectors, or other relic relativistic particles)
  while the sterile state, which participates further
in oscillations with the electron neutrino is initially empty. In that case
the two effects will
be simply additive, and the uppermost long-dashed curve in Figs.2 presents the
overproduction of $^4$He for such specific situation.

\ \\

\hspace{2cm}\includegraphics[scale=0.4]{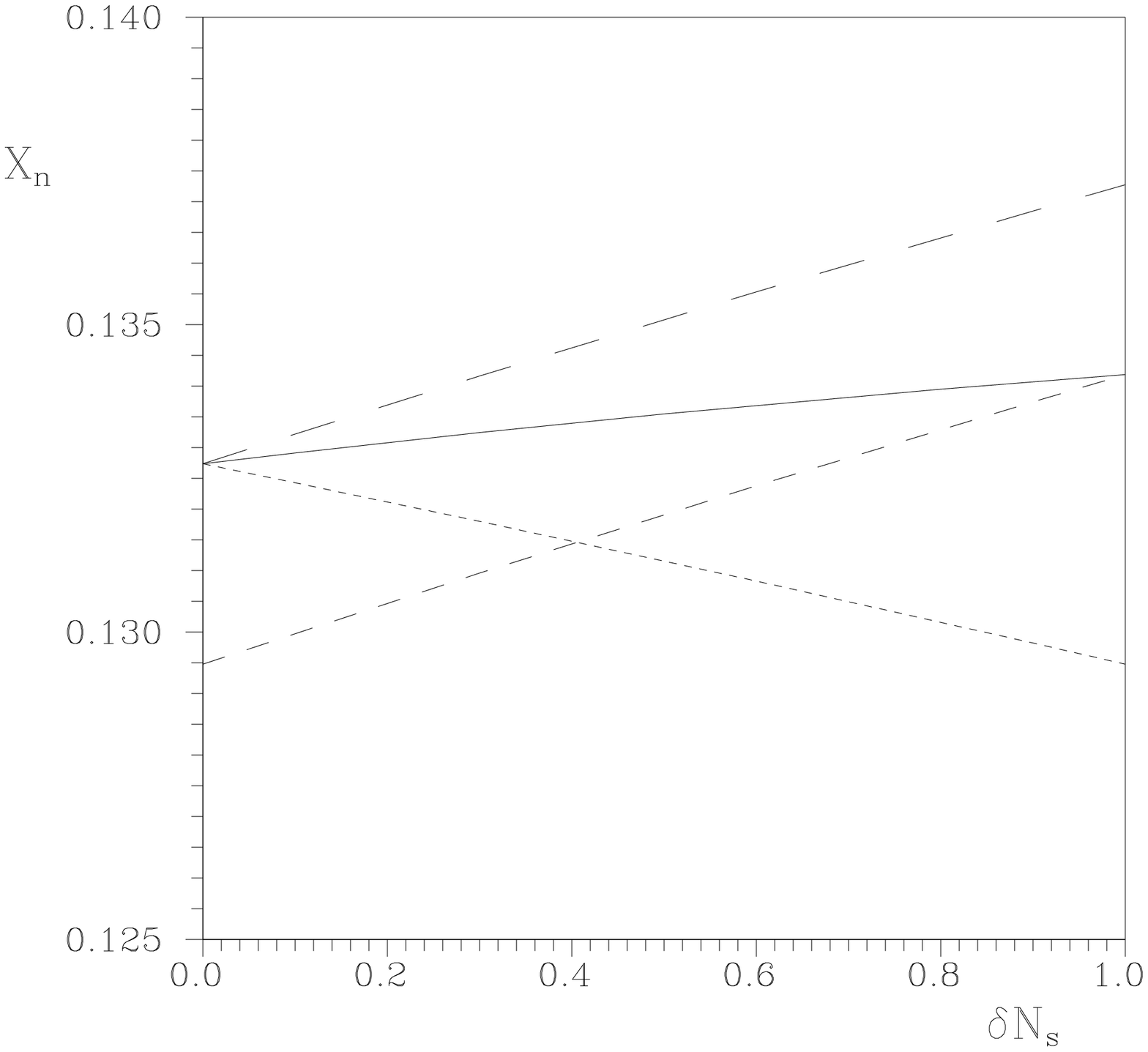}
\ \\
{\footnotesize Fig.2b. The  solid curve presents the frozen
neutron number density relative to
nucleons $X_n^f=n_n^f/n_{nuc}$ as a function  of
the sterile neutrino initial population, at $\delta m=\pm 10^{-9}$
eV$^2$, $\sin^2 2\theta=1$. The dotted curve presents
the
kinetic effect, while the lower dashed curve presents
energy
density increase effect. The uppermost long dashed curve corresponds to the
total effect  if  the effects were simply additive.}

However, due to the fact that $\delta N_{kin}$ is a decreasing function of
$\delta N_s$,
   naively adding the two effects  exaggerates  $Y_p$ overproduction,
and hence would
define stronger bounds than the real ones.

{\bf Second: The direction of  the shift  of the constraints  should be as follows:
in case $\delta N^0_{kin}\delta N_s<\delta N_s$ the constraints will
be strengthened in comparison with the $\delta N_s=0$ constraints, while
in the opposite case they will be relaxed.}

In fig.2a   the solid curve, presenting the total effect, is a decreasing function of
 $\delta N_s$ because $\delta N^0_{kin}>1$, i.e. for that set of oscillation parameters the overproduction of $^4$He due to oscillations decreases with
 the increase of the initial population of the sterile neutrino state. Obviously
 the suppression effect (ii) dominates. In that case we expect relaxation of the
 cosmological constraints compared to the case of initially zero  population of $\nu_s$.

 Fig.2b  presents the results for a set of oscillation parameters for which
  $\delta N^0_{kin}<1$, then as it is illustrated the total effect is an increasing function of $\delta N_s$, the dynamical effect (i) dominates over (ii),  thus the overproduction
 of $^4$He increases in respect to the case of $\delta N_s=0$ and hence, the cosmological constraints  must become more stringent   with the increase of $\delta N_s$.
 And as far as $\delta N^0_{kin}=\delta N_{tot}$  at  $\delta N_s=0$,
 the BBN constraints,   corresponding to  $^4$He observational uncertainty expressed as  $\delta N_{tot}<1$,  will
be strengthened.
For example, $^4$He uncertainty $\delta Y_p \sim 0.007$ corresponds to   $\delta N_{tot} \sim 0.54$. Thus  at  $\delta N_s=0$ $\delta N^0_{kin}\sim 0.54<1$,  and hence  strengthening of the cosmological constraints is expected.

 The exact form of the  cosmological constraints obtained by a detail numerical  study of the effects (i) and (ii) and corresponding
to such  $^4$He overproduction and different $\delta N_s$ values is presented in the next
section.

In case of bigger  $^4$He  uncertainty, parametrized by    $\delta N_{tot}>1$,
the term (ii) dominates (as illustrated in Fig.2a), leading to a decrease of the $^4$He overproduction due to oscillations
 and hence, relaxation of the cosmological constraints should be expected.~\footnote{In more detail this case will be discussed in a following paper~\cite{markir}.}
It is interesting to note that contrary to some prejudice,
even for $^4$He uncertainty equivalent to $\delta N_{tot}>1$ cosmological constraints
on oscillation parameters still persist, provided that $\delta N_s<1$ and that a proper description of the  neutrino energy spectrum distortion is made.

\section{BBN constraints for partially filled $\nu_s$}

For calculating cosmological  constraints on oscillation
parameters, I have followed the kinetic approach described in
detail in ref.~\cite{MPD}. The case of
$\nu_e\leftrightarrow \nu_s$ oscillations effective after electron
neutrino decoupling is considered. For simplicity mixing just in the electron
sector is assumed.
 The more general case of a sterile neutrino state being partially filled before the
oscillations become effective is studied.
In a previous work~\cite{MPD} it was shown that for a wide range of  $\delta N_s$ values the kinetic effects of
oscillations, namely spectrum distortion of the electron neutrino and
lepton asymmetry growth,  play a considerable
role, hence  rough analytical estimations of the
kinetic effects are not applicable~\cite{comment, nashi}. Therefore, here
 an exact numerical analysis of $\delta N_s$ effects (i) and
(ii)  on pre-BBN nucleons freezing and
the production of $^4$He in BBN with oscillations is provided.

The kinetic equations for the density matrix of neutrino and antineutrino
ensembles are simultaneously and selfconsistently  solved with the kinetic equations governing
the evolutions of nucleons from the period of neutrino decoupling  $T \sim 2$
MeV till the nucleons freezing epoch, down to $T \sim 0.3$ MeV.
The neutrino spectrum is described by 1000 bins for the nonresonant case, and by
more than 5000 bins (depending on the asymmetry growth region of oscillation
parameters).
This detail study, performed for a wider range of parameters,
confirms the importance of the kinetic effect at non-zero $\delta N_s$, found  in
ref.~\cite{MPD}.
I  calculate primordial $^4$He yield
 $Y_p(\delta N_s, \delta m^2,sin^22\vartheta)$ at
 different  $\delta N_s$ values   for the full
set of oscillations parameters of the model:
for all mixing angles
$\vartheta$ and
 mass differences
$\delta m^2 \le 10^{-7}$ eV$^2$.

For the analysis of the constraints  the observational
 uncertainty of the primordially produced   $^4$He  is
assumed $\delta Y_p < 0.007$ in correspondence with the accepted systematic error in the
 $^4\!$He measurements~\cite{He}.
Then the maximum  possible value of $\delta N_s$  at BBN
epoch is constrained on the
basis of BBN considerations:
Using the approximate empirical formula
$\delta Y_p \sim 0.013 \delta N_{tot}$, $\delta Y_p<0.007$  corresponds to
$\delta N_s < 0.54$. So, in our analysis we have varied $\delta N_s$
in the range
$0.0\le\delta N_s\le 0.5$ with a step 0.1. The case $\delta N_s>0.54$
corresponds to
higher  $^4\!$He uncertainty and will not be studied  here.

This choice of maximum $\delta N_s$ is also supported by the results of the
recent
standard BBN analysis + $^4\!$He observational data  with an input the baryon
density value from WMAP data, which
provide bounds on the number of
additional neutrino species at BBN  in the range
$\delta N_s<0.1 - 0.5$~\cite{cuoco,cyburt,barger,han,crotty}.


Uncertainty $\delta Y_p=0.007$ corresponds to $\delta Y_p/Y_p \sim 3\%$,
so in this analysis, I have assumed that primordial $^4$He abundance is known
with an accuracy better than
$3\%$, and will call  the isohelium contours corresponding to  $\delta Y_p=0.007$  '3\% $^4$He  overproduction contours', and the obtained BBN constraints '3\% $^4$He constraints'.


 In Fig.~2 I present the calculated
cosmological constraints on oscillation parameters  for the $\delta N_s=0.1$,
and  $\delta N_s=0.5$ for the
resonant (to the left) and the non-resonant
(to the right) oscillation cases. The region upwards of the corresponding curves is
the cosmologically excluded. The upper curve is for  $\delta N_s=0.1$ case, the lower one -- for
$\delta N_s=0.5$.
$\delta N_s=0$ constraints are  given for comparison by the dashed contours.

\mbox{\hspace{1cm}}\epsfig{figure=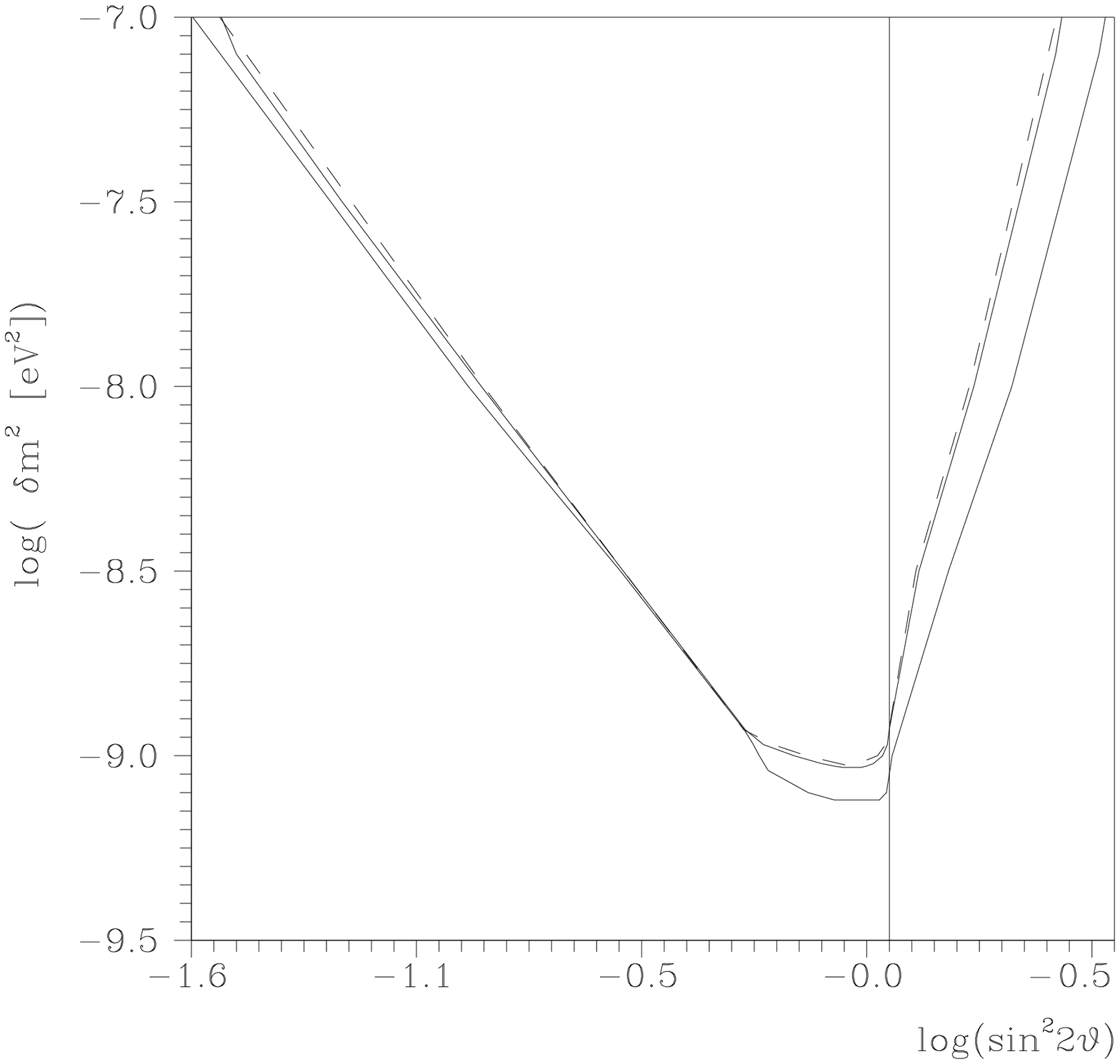,width=10cm}\\
{\bf Figure 3: }{\small BBN constraints on oscillation parameters
for the resonant
(l.h.s.) and the non-resonant $\nu_e\leftrightarrow\nu_s$ oscillations
and for initial
degrees of population of the sterile neutrino state $\delta N_s=0.1$ and
$\delta
N_s=0.5$. The dashed contours present the constraints for  $\delta N_s=0$ case
for comparison.}
\ \\

The analysis  shows that the two effects (i) and (ii) of  $\delta N_s$, nearly
compensate each other for small $\delta N_s$ values.
Hence, the cosmological constraints for $\delta N_s=0.1$ slightly differ
 from the ones for  $\delta N_s=0$, as illustrated in  Fig.3.
 For $\delta N_s>0.1$ $\delta Y_d$ dominates over the suppression term $\delta Y^s_k$,
 so the constraints are strengthened.
As a whole the cosmological constraints  in the resonant case are
slightly  changed compared to the case with initially zero
sterile state,
while in the non-resonant case the
change is more noticeable, and the constraints are becoming more stringent with
the increase of  $\delta N_s$.

The precise account of the kinetic effects of oscillations on BBN,
i.e. the study of the distorted spectrum distribution of neutrinos and its effect on nucleons kinetics,
allows strengthening the cosmological constraints by an order of magnitude towards smaller mass differences in comparison with calculations considering just the integral effect like neutrino number density depletion.


\section{Discussion}

The presence of a {\it  non-empty} sterile state before
$\nu_e\leftrightarrow \nu_s$  oscillations was not considered in
previous cosmological constraints on oscillation parameters.
This work is a  step towards generalizing  the
cosmological constraints on oscillation parameters of neutrino.
It discusses the effect  of partially filled sterile state before oscillations
on primordial production of $^4$He and on the BBN  oscillations constraints.

It is shown that initially non-zero $\nu_s$ state has two fold effect on $^4$He:
on one hand it
affects Universe dynamics leading to overproduction of $Y_p$, while on the other hand
its presence suppresses oscillations kinetic effects causing a decrease of $Y_p$ overproduction.
Thus, depending which effect dominates, non-zero initially $\nu_s$ may cause strengthening or relaxation of the cosmological constraints on oscillation parameters.

I have calculated
the isohelium contours corresponding to  $3\%$ overproduction of  $^4\!$He,
in a model of BBN with electron--sterile oscillations effective after neutrino
decoupling and for the general case of the sterile neutrino state being partially
populated before oscillations, $0.0<\delta N_s\le 0.54$.
The constraints become more stringent  with the increase of
  $\delta N_s$ value. So, the presence of non-zero initial population
 $\delta N_s\le 0.54$ strengthens the  $3\%$ $^4\!$He cosmological constraints.

 These cosmological constraints  are obtained for the case of 2-neutrino mixing.
Further generalization of the constraints includes the  account of mixing between active
neutrinos,  which  has been proved
 important in the resonant oscillation case for
oscillations proceeding before neutrino decoupling~\cite{Villante}.

The constraints are obtained for the natural assumption  that at BBN epoch the initial lepton
asymmetry is
of the order of the baryon one. However, in fact  the lepton asymmetry in
the neutrino
sector is not strongly constrained~\cite{pastor}. We expect that,
provided  a small lepton asymmetry $L<<0.1$ is present, it may be large enough
to relax or alleviate the discussed BBN bounds on neutrino oscillations.
For $\delta N_s=0$ case it was proven that
larger than $10^{-7}$ lepton asymmetry may strongly suppress oscillations effective after neutrino decoupling and change the constraints,
while initial lepton asymmetry larger than $10^{-5} -- 10^{-4}$ can
 alleviate them~\cite{NPB98,NPS}.

Cosmological constraints can be relaxed also if the systematic error of $Y_p$ is  higher than 0.007. However, it is interesting to note that even for $\delta Y_p \sim 0.01$,  and  a considerable initial population of $\nu_s$ $\delta N_s<1$,
the constraints  may not be removed, but just
relaxed, as the results of ref.~\cite{markir} suggest.

\ \\
\ \\

I thank M. Chizhov, A.Dolgov, J.-M. Frere, R. Mohapatra, A. Smirnov and M. Tytgat  for interesting discussions.
I am grateful to the organizers of the  Workshop on Neutrino and the Early Universe, Trento 2004,
M. Lindner and G. Raffelt  for the
possibility to present and discuss the preliminary results of this work at the workshop.
I appreciate the
Regular  Associateship  at the Abdus Salam ICTP, Trieste.
The numerical analysis was provided also thanks to CERN computing facilities.
This work  was supported in part by Belgian Federal Government
(Federal Public Planning Service Science Policy (PPS Science Policy)).

\end{document}